\documentclass[twocolumn,showpacs,prl]{revtex4}

\usepackage{graphicx}
\usepackage{amssymb}

\begin{document}

\title{The Boson Peak and its Relation with Acoustic Attenuation in Glasses}

\author{B. Ruffl\'e$^1$}
\author{D.A. Parshin$^{1,2}$}
\author{E. Courtens$^1$}
\author{R. Vacher$^1$}
\affiliation{$^1$Laboratoire des Collo\"{\i}des, Verres et Nanomat\'eriaux, UMR 5587 CNRS\\
Universit\'e Montpellier II, F-34095 Montpellier Cedex 5, France\\
$^2$Saint Petersburg State Technical University, 195251 Saint Petersburg, Russia}
\date{Received 11 july 2007; revised 31 august 2007}

\begin{abstract}
Experimental results on the density of states and on the acoustic modes of glasses in the THz region are compared to the predictions of two categories of models. A recent one, solely based on an elastic instability, does not account for most observations. Good agreement without adjustable parameters is obtained with models including the existence of non-acoustic vibrational modes at THz frequency, providing in many cases a comprehensive picture for a range of glass anomalies.
\end{abstract}

\pacs{63.50.+x, 78.35.+c, 81.05.Kf}

\maketitle

The boson peak is an excess in the vibrational density of states (VDOS), $g(\omega )$, observed in many glasses at frequencies $\omega /2 \pi $ of the order of one THz. It appears as a hump in $g(\omega )/\omega ^2$ {\em vs.} $\omega $, above the acoustic Debye level $g_{\rm D}(\omega )/\omega ^2$. It is typically located at $\Omega _{\rm BP} \sim 0.1 \times \omega _{\rm D}$, where $\omega _{\rm D} $ is the Debye frequency. This excess produces the well-known specific heat anomaly of glasses at temperatures $T \simeq \hbar \Omega _{\rm BP}/5 k_{\rm B} \sim  10 $ K \cite{Zel71}. It is generally agreed that the boson peak is a vibrational signature of the disordered structure of glasses beyond the nanometer scale. Its correct understanding is thus of considerable importance. Two main categories of dynamical boson-peak models currently exist. The first, which we call {\em harmonic random matrix} (HRM) models, is based on the concept that purely harmonic elastic disorder produces an excess of low frequency modes. The alternate picture is that there exist in glasses additional --non-acoustic-- {\em quasi-local vibrations} (QLVs) at low frequencies \cite{Lif43,Kar83}. The purpose of the present Letter is to compare HRM and QLV models to actual experimental results. We find that QLVs, using only independently determined parameters, mostly provide a much better agreement between theory and experiment than HRMs can do.

The HRM models postulate randomly fluctuating spring constants $K_{ij}$, as {\em e.g.} in \cite{Sch98,Tar01,Kan01,Gri02,Gur05}. These assume a distribution $p(K)$, generally bounded between two values $K_{\rm min} < K_{\rm max}$. A truncated gaussian distribution centered at $K_0 > 0$ but extending to $K_{\rm min} < 0$ is used in \cite{Sch98}. The system becomes mechanically unstable if $- K_{\rm min} $ is above a threshold. Just before this instability, a low frequency excess in the VDOS appears as precursor, which is interpreted as the boson peak \cite{Sch98}. In \cite{Tar01}, a square distribution $p(K)$ is used instead. A negative $K_{\rm min}$ is not necessary if $p(K)\propto 1/K$ over the interval $\{ K_{\rm min} , K_{\rm max}\}$ as shown in \cite{Kan01}. Such a distribution can be rationalized by free-volume considerations. A similar distribution, including some negative force constants, was used in \cite{Gri02} and handled with so-called Euclidean Random Matrices. The shape and position of the predicted boson peaks are not universal as they depend on the selected $K$-distribution. While earlier treatments were based on microscopic harmonic models, more recently macroscopic tensorial elastic approaches were developed \cite{Gur05,Sch07}. In \cite{Sch07}, the gaussian disorder affects only the shear modulus which tends to zero at critical disorder. This now allows performing realistic comparisons with experimental results, and it will be used below. In all these models, the boson peak is produced by shifting the maximum in $g(\omega )/\omega ^2$ to low $\omega $ owing to a softening of the elastic constant distribution.

On the other hand, inelastic neutron-scattering measurements of the dynamic structure factor of silica showed unambiguously that its boson-peak modes are not sound waves \cite{Buc86}, providing an early justification for the QLV hypothesis. That the boson peak of this network glass relates to librations of structural units has since been confirmed by hyper-Raman scattering \cite{Heh00}, and similarly for boron oxide \cite{Sim06}. Additional manifestations of excess excitations are the thermal and acoustic anomalies observed below liquid-He temperatures that are described by two-level systems \cite{Phi81}. A theory encompassing these and QLVs is the soft potential model \cite{Kar83,Par94}. The latter predicts an onset in the {\em excess} VDOS, $g_{\rm V} = g - g_{\rm D}$, with $g_{\rm V} / \omega  ^2 \propto \omega ^2$ \cite{Par94}, evolving into a boson peak when the total number of QLVs is appropriately limited \cite{Ram93}. This is now further understood on the basis of a physical model that considers the {\em mechanical instability} of an initially uniform distribution of quasi-harmonic QLVs of density $g_0(\omega )$, interacting via their common strain field \cite{Gur03,Par07}. The authors assume that $g_0(\omega )$ has a high cut-off, at $\omega _0 \sim \omega _{\rm D}$. The interaction being weak, it only destabilizes those oscillators of relatively low frequencies, below $\omega _{\rm c} \ll \omega _0$ \cite{foot1}. The destabilized oscillators, restricted by their anharmonicity, form a renormalized density $g_{\rm c}(\omega ) \propto \omega $ up to $\omega _{\rm c} $ \cite{Gur03}. This result is independent from the initial $g_0(\omega )$ as well as from the size of the anharmonicity. A last but crucial step is that, owing to anharmonicity, the displaced oscillators produce random static forces affecting other nearby oscillators. This creates a soft gap, $g_{\rm V}(\omega ) \propto \omega ^4$, up to $\omega _{\rm b} < \omega _{\rm c}$. The reader is referred to \cite{Gur03,Par07} for details. This model thus leads to a peak in $g_{\rm V}(\omega )/\omega ^2$, with an onset in $\omega ^2$ up to $\simeq \omega _{\rm b} $, and a decay in $\approx \omega ^{-1}$ between $\omega _{\rm b}$ and $ \omega _{\rm c}$. While the onset power is universal, the decay above $\omega _{\rm b}$ depends on the separation between $\omega _{\rm b}$ and $ \omega _{\rm c}$. For small separation, the decay is between $\omega ^{-1}$ and $\omega ^{-2}$ \cite{Par07}, in good agreement with observation \cite{Buc07}. Contrary to HRMs, the QLV boson peak is {\em not} a shifted down end of acoustic branches. The high frequency VDOS is but slightly modified by the successive interactions. Complemented with the soft-potential model, the QLV model makes specific predictions concerning high-frequency acoustic modes and their Ioffe-Regel limit beyond which they cease existing as plane waves \cite{Par07,Par01}. Those will be checked below.

\begin{figure}
\includegraphics[width=8.5cm]{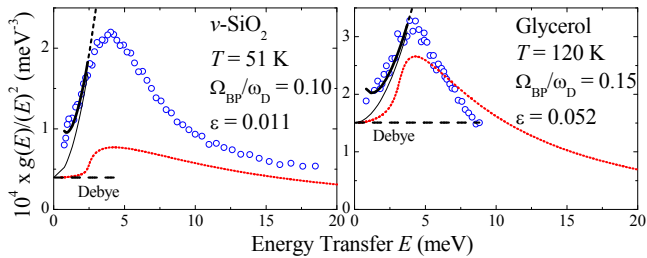}
\caption{The measured boson peaks of $v$-SiO$_2$ \cite{Wis98} and glycerol \cite{Wut95} compared to the Debye levels (dashed horizontal lines), to the best prediction of the HRM model \cite{Sch07} (dotted lines), and to the onset calculated using the soft-potential model, Eq. (1a) (thin solid lines) and Eqs. (1a+b) (thicker solid lines). (color online)} 
\end{figure}

First consider the VDOS in the boson peak region. High quality data are available on many glasses, in particular from neutron scattering. For example, the experimental results, normalized to unity and divided by $E^2$, where $E = \hbar \omega $, are shown in Fig. 1 for the strong network glass silica, $v$-SiO$_2$ \cite{Wis98}, and for the intermediate molecular one glycerol \cite{Wut95}. The corresponding Debye levels, $g_{\rm D} (E)/E^2$, are shown by dashed lines. These are calculated from the known velocities of longitudinal (LA) and transverse (TA) acoustic modes, $v_{\rm L}$ and $v_{\rm T}$ respectively. The boson peak of $v$-SiO$_2$, at $E_{\rm BP} = \hbar \Omega _{\rm BP} \simeq 4$ meV, exhibits the well known large relative excess $g_{\rm V}/g_{\rm D} \sim 5$. In glycerol, this ratio is much smaller, $\sim 1.2$. The dotted lines in Fig. 1 illustrate for comparison the predictions of the HRM model, Eq. (2) of \cite{Sch07}. These are determined iteratively using the {\em experimental} values of the sound velocities and of $\Omega _{\rm BP}/ \omega _{\rm D}$ to extract the appropriate bare velocities and separation parameter $\varepsilon $ of \cite{Sch07}. It is obvious from Fig. 1 that the elastic instability model falls short of reproducing the observed peak strength. This already suggests that additional modes should be involved. For further comparison, the prediction of soft potentials is calculated below $E_{\rm BP}$, in the region where the growth of $g_{\rm V} / E^2$ is in $E^2$, using \cite{Ram93,Buc07}
$$ g_{\rm V}(E) / E^2 = \frac{P_{\rm s}}{24} \; \frac {1}{W^3} \; \left ( \frac{E}{W} \right ) ^2 \;\;\; . \eqno{(1a)} $$
Here $P_{\rm s}$ is the density of additional modes around the eigenvalue zero, calculated per atom, and $W$ is the crossover energy between vibrational and tunneling states. These parameters are obtained from thermodynamical data around liquid-helium temperature ($T$), and are available, both for silica \cite{Ram97} and for glycerol \cite{Tal01}. Using (1a), and adding the Debye contribution $g_{\rm D} (E)/E^2$, one obtains the thin lines shown in Fig. 1. It is well known that there exist, in addition to the VDOS, quasi-elastic contributions to the scattering signal arising from relaxations in double-well potentials. This is described with the same soft potential parameters \cite{Buc07},
$$g_{\rm rel}(E,T)/E^2 \approx \frac{P_{\rm s}}{2} \; \frac {1}{E W^2} \; \left ( \frac{k_{\rm B}T}{W} \right ) ^{3/4} \;\;\; , \eqno{(1b)} $$
where $k_{\rm B}T$ is the thermal energy. Adding $g_{\rm rel} $ to $g_{\rm V} $, a quantitative agreement is obtained in the onset region below $E_{\rm BP}$, shown by thick lines in Fig. 1. This is not a trivial result, as the parameters $P_{\rm s}$ and $W$ entering Eqs.(1) are derived from independent very low $T$ measurements. It emphasizes the self-consistency of QLVs and two-level systems embodied in the soft potential model. It is an additional indication that the excess indeed arises from QLVs. The solid lines in Fig. 1 do not account for the saturation in the growth of $g(E)/E^2$ near $E_{\rm BP}$. This could be easily included, as shown {\em e.g.} in \cite{Buc07}. According to \cite{Ram93,Par07}, it requires {\em at least} one additional parameter which is beyond the purpose of the present discussion.

\begin{figure}
\includegraphics[width=8.5cm]{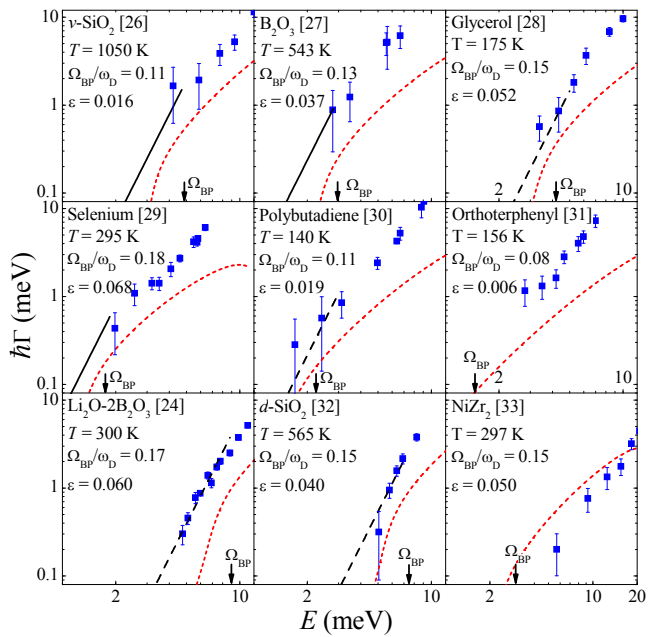}
\caption{The full widths $\hbar \Gamma $ of x-ray Brillouin spectra measured on nine glasses in function of the mode energy $E$, both derived from adjustments with damped harmonic oscillator lineshapes \cite{Del98, Mat01, Set98, Sco04, Fio99, Mon98, Ruf06, Ruf03, Sco06}. The dotted curves are calculated following \cite{Sch07} as explained in the text, leading to the parameter $\varepsilon $ which is indicated. The lines of slope 4 are either calculated using Eq. (2) (solid lines), or traced through the experimental points (dashed lines), terminating then at the $E_{\rm IR}$ values obtained from the data as explained in \cite{Ruf06}. Arrows indicate the approximate boson-peak positions taken from \cite{Ruf07} where the experimental $\Omega _{\rm IR}$ values are also discussed.
(color online)}
\end{figure}

A second important information is the linewidth of acoustic modes of very high frequency $\Omega /2\pi $, which near the THz range should increase as $\Gamma \propto \Omega ^4$, both in the HRM and QLV models. In HRMs, the acoustic width arises from elastic Rayleigh scattering by disorder. With soft potentials, it is the resonant absorption of sound by excess modes that gives the leading contribution \cite{Ram97},
$$ \hbar \Gamma  = \frac {\pi }{8} \; C_{\rm L} \; E \left ( \frac{E}{W} \right ) ^3 \;\;\; . \eqno{(2)} $$
Here, $\Gamma $ is the full width of LA modes relating to the {\em energy} mean free path $\ell = v_{\rm L}/\Gamma$, $E$ is the phonon energy $\hbar \Omega $, and $C_{\rm L}$ is the tunneling strength for LA modes that can be obtained from the height of the acoustic attenuation plateau near liquid-He $T$. This strong growth, $\Gamma \propto E^4$, leads to a rapid end of acoustic plane waves as $E $ increases. This occurs at a Ioffe-Regel crossover $E_{\rm IR} = \hbar \Omega _{\rm IR}$ where $\ell $ decreases down to half the wavelength, {\em i.e.} for $\Gamma = \Gamma _{\rm IR} = \Omega _{\rm IR} / \pi $ \cite{Ruf06}. This crossover happens to fall at frequencies and wavevectors well above those reached in optical or UV Brillouin scattering, but most often near the lowest values reachable with inelastic x-ray scattering. The energy $E_{\rm IR}$ can be determined experimentally provided $\Gamma (E)$ is available over an adequate range \cite{Ruf06}. It can also be calculated from Eq. (2) and the known soft-potential parameters, $E_{\rm IR} = 2(\pi ^2 C_{\rm L})^{-1/3}W$. In the few cases where both can be done, their agreement is remarkable \cite{foot_IR}.

The current situation for nine glasses is summarized in Fig. 2. The dots show the x-ray data \cite{Del98, Mat01, Set98, Sco04, Fio99, Mon98, Ruf06, Ruf03, Sco06}. These are full widths of damped harmonic oscillator lineshapes adjusted to x-ray Brillouin spectra after convolution with the instrumental response. The HRM predictions (dotted lines) are calculated using Eq. (4) of \cite{Sch07}. Again the known values of $v_{\rm L}, v_{\rm T}$, and $\Omega _{\rm BP}/ \omega _{\rm D}$ are used to extract the separation parameter $\varepsilon $. It is clear that in all but one case the best HRM predictions fall well below the observed widths. Interestingly, the single exception is a metallic glass \cite{Sco06} for which a real Ioffe-Regel crossover was not observed \cite{Cou07} and might not exist. For comparison, the QLV predictions below the crossover are drawn as solid lines for the three cases for which reliable values of {\em both} $W$ and $C_{\rm L}$ are available \cite{foot2}. Above the crossover, Eq. (2) ceases to apply, and the widths cease increasing in $E^4$. This presentation shows that the experimental widths smoothly prolongate the QLV prediction, which again supports the quantitative validity of this model on the basis of independently determined parameters. However, it also illustrates that the x-ray data unfortunately starts at somewhat too high energy in these three glasses to properly investigate the onset region $\hbar \Gamma \propto E ^4$. On the opposite, this onset falls at sufficiently high $E$ to be well observed in lithium diborate Li$_2$O-2B$_2$O$_3$ \cite{Ruf06} and in densified silica glass of density 2.6 g/cm$^3$, $d$-SiO$_2$ \cite{Ruf03}. In these cases, as well as for the four other glasses in Fig. 2, either one or both soft-potential parameters are unknown. For glycerol and polybutadiene $W$ is known, but for $C$ one only has an average value derived from low $T$ thermal conductivity. It turns out that $C_{\rm L}$ and $C_{\rm T}$ (for TA modes) can differ by factors as large as $\sim 3$. Thus, the solid line in Fig. 2 cannot be drawn for these two glasses. Returning to the HRM predictions, it is also remarkable that the slope of the dotted lines tapers off gently in all cases, and that no single line crosses the value $\Gamma = \Omega / \pi $ within the ranges shown. This means that the HRM model does not predict Ioffe-Regel crossovers at places where these are in fact observed \cite{Ruf07}.

\begin{figure}
\includegraphics[width=8.5cm]{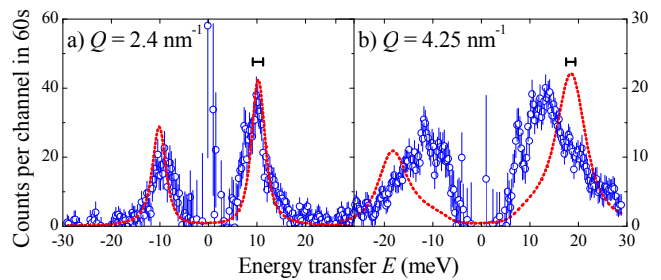}
\caption{Two inelastic spectra obtained on lithium diborate \cite{Ruf06} at wavevectors $Q$ near and above the Ioffe-Regel crossover at $Q_{\rm IR} \simeq 2.1$ nm$^{-1}$. The dotted curves are the best predictions of the HRM model \cite{Sch07}. The relative amplitudes are significant. The instrumental full widths at half maximum for the Brillouin peaks, including the effect of the finite collection angle, are indicated by horizontal bars. (color online)} 
\end{figure}

It is also of interest to compare spectral shapes with specific HRM predictions of \cite{Sch07}. This is illustrated in Fig. 3 for spectra observed on Li$_2$O-2B$_2$O$_3$ \cite{Ruf06}. In drawing Fig. 3, the HRM lineshapes have been convoluted with the instrumental response function, and their height was adjusted to the corresponding experimental peak amplitude at low $Q < Q_{\rm IR}$. The instrumental response being wider than the spectral broadening, the latter is hard to judge just inspecting Fig. 3a. Only the numerical analysis of the data leads to the widths given in the corresponding panel in Fig. 2. However, as $\Omega $ increases beyond $\Omega _{\rm IR}$, the HRM predictions, {\em including the peak position}, start departing strongly from the observed spectra which also become very wide. This presentation clearly shows that the {\em best} HRM prediction cannot provide an explanation for the observed spectra. In the QLV model, one expects that in the crossover region the acoustic modes hybridize with the boson-peak modes. In such a strong coupling case, perturbation approaches become inadequate. For this reason, a reliable expression for the Brillouin lineshapes in the QLV model is not yet available.

In conclusion, it clearly appears that acoustic modes alone are not able to account for observations on most glasses in the THz range. The acoustic modes are merely a subset of all the low frequency vibrations in disordered systems. QLVs predicted by the soft potential model take these into account using parameters that are determined fully independently by two-level-system measurements \cite{Phi81,Par94}. It is truly remarkable that the latter, performed at low frequencies and sub-liquid-He temperatures, agree so well with hypersonic measurements at elevated temperatures. For network glasses, there exists separate evidence on the non-acoustic nature of the boson-peak excess which relates to rigid librations of structural units \cite{Buc86,Heh00,Sim06}. For molecular glasses, the disagreement between the observed widths and the HRM predictions in Fig. 2 is also striking. It is intuitive that for these the lowest frequency non-acoustic modes could be rigid librations as well, though there exists so far little experimental evidence for this. Contrary to crystals in which modes are orthogonal, in disordered systems the QLVs can linearly couple to strains and to planar acoustic modes which are not true eigenmodes. For this reason the acoustic modes seen in Brillouin scattering are strongly affected as their frequency nears the boson peak. The observed relation between  $E_{\rm IR}$ and  $E_{\rm BP}$ \cite{Par01,Ruf06,Ruf07} is also well explained by this coupling, as shown in \cite{Par07}.


\begin{thebibliography}{99}
\bibitem{foot_IR}For $v$-SiO$_2$, B$_2$O$_3$, and Se, we find experimentally for $E_{\rm IR}$ the values $5.2 \pm 1.7$, $3.0\pm 0.9$, and $2.4\pm 0.5$ meV, in that order. From the soft-potential parameters one calculates 4.7, 2.8, and 1.9 meV, with unknown error bars.
\end{thebibliography}
\end{document}